\documentclass[cernpreprint, atlasdraft=false, UKenglish, texmf, orcidlogo]{atlasdoc}
 
\usepackage{atlaspackage}
\usepackage{atlasbiblatex}

\usepackage{atlasphysics}
 
\usepackage{letltxmacro}
\LetLtxMacro{\oldcite}{\cite}
\renewcommand{\cite}[1]{\mbox{\oldcite{#1}}}
 
\usepackage{multirow}
 
\usepackage{siunitx}
 
\usepackage{calc}  
\usepackage{array} 
\newlength\mylen
\graphicspath{{logos/}{figures/}}
 
\usepackage{ANA-HIGG-2022-12-PAPER-defs}

\AtlasTitle{Measurement of the $H \to \gamma \gamma$ and $H \to ZZ^* \to 4 \ell$ cross-sections in $pp$ collisions at $\sqrt{s}=13.6$ TeV with the ATLAS detector}

\AtlasAbstract{
The inclusive Higgs boson production cross-section is measured in the di-photon and the $ZZ^* \to 4 \ell$ decay channels
using 31.4 and 29.0~\ifb\ of $pp$ collision data respectively, collected with the ATLAS detector at a centre-of-mass energy of $\sqrt{s}=13.6$~\TeV.
To reduce the model dependence, the measurement in each channel
is restricted to a particle-level phase space that closely matches the channel's detector-level kinematic selection, and it is
corrected for detector effects. These measured fiducial cross-sections are \FiducialYYShort,
and \FiducialZZShort,
in agreement with the corresponding Standard Model predictions of \FiducialYYExpNum\ and \FiducialZZExpNum.
Assuming Standard Model acceptances and branching fractions for the two channels, the fiducial measurements are extrapolated to the full phase space
yielding total cross-sections of \TotalYYShort\ and \TotalZZShortNum\ at $13.6$~\TeV\ from the
di-photon and $ZZ^* \to 4 \ell$ measurements respectively. The two measurements are combined into a total cross-section
measurement of \TotalCOMBShort, to be compared with the Standard Model prediction of \TotalCOMBExp.
}

\AtlasRefCode{HIGG-2022-12}
 
\PreprintIdNumber{CERN-EP-2023-114}

\AtlasJournalRef{Eur. Phys. J. C 84 (2024) 78}
\AtlasDOI{10.1140/epjc/s10052-023-12130-5}

\hypersetup{pdftitle={ATLAS document},pdfauthor={The ATLAS Collaboration}}
 
\begin{document}
 
\maketitle
 
\tableofcontents

\section{Introduction}
\label{sec:intro}

Since the discovery of the Higgs boson (denoted by $H$ in this paper) by the ATLAS and CMS
collaborations~\cite{HIGG-2012-27,CMS-HIG-12-028} at
the Large Hadron Collider (LHC)~\cite{Evans:2008zzb}, a major goal  of the physics programme of the LHC experiments
has been to measure its properties and determine whether they correspond to those predicted by the Standard Model
(SM) of particle physics 
or involve new phenomena beyond those described by this theory.
 
Despite the small branching ratios of Higgs boson decays into two photons and four leptons ($ZZ^* \rightarrow 4\ell$, with $\ell=$ e or $\mu$),
these decay channels are particularly interesting for this purpose given the excellent
mass reconstruction and photon and lepton identification efficiencies of the ATLAS detector.

In 2022 the LHC increased the centre-of-mass energy of the proton--proton (\pp) collisions to the new world-record value of 13.6~\TeV.
In the analysis described in this paper, 31.4 and 29.0~\ifb\ of \pp collision data recorded with the ATLAS detector are used to derive
the first ATLAS measurements of the \hgg and  \hfourl\ cross-sections at this new energy.
The cross-sections are measured inclusively in the production modes. They are derived via a fit to the reconstructed invariant mass spectra.
To reduce the model dependence, the measurements are restricted to a kinematic phase space of the Higgs boson decay
products defined at particle level that closely matches the selection criteria applied at the detector level, and they are
corrected for detector effects (unfolding). This approach is usually referred to as a ‘fiducial cross-section’ measurement.
The fiducial cross-sections of the two decay channels are also extrapolated to the full phase space, corrected for the predicted SM decay branching ratios
and combined.
The ATLAS and CMS experiments have previously published fiducial cross-section measurements for the \hgg and \hfourl\
processes at different centre-of-mass energies up to 13 \TeV~\cite{HIGG-2013-10,HIGG-2018-29,HIGG-2019-13,HIGG-2013-22,CMS-HIG-14-016,CMS-HIG-19-016,CMS-HIG-19-001,CMS-HIG-14-028,CMS:2023gjz}. These results do not show any significant deviation from the SM predictions.
 
All results reported in this paper assume a value of the Higgs boson mass of $125.09 \pm 0.24$ \GeV~\cite{HIGG-2014-14}.


\section{ATLAS detector}
\label{sec:detector}

\newcommand{\AtlasCoordFootnote}{
ATLAS uses a right-handed coordinate system with its origin at the nominal interaction point (IP)
in the centre of the detector and the \(z\)-axis along the beam pipe.
The \(x\)-axis points from the IP to the centre of the LHC ring,
and the \(y\)-axis points upwards.
Polar coordinates \((r,\phi)\) are used in the transverse plane,
\(\phi\) being the azimuthal angle around the \(z\)-axis.
The pseudorapidity is defined in terms of the polar angle \(\theta\) as \(\eta = -\ln \tan(\theta/2)\).
Angular distance is measured in units of \(\Delta R \equiv \sqrt{(\Delta\eta)^{2} + (\Delta\phi)^{2}}\).}
 
\label{sec:atlas3a}
 
The ATLAS experiment~\cite{PERF-2007-01,Run3DetPaper} at the LHC is a multipurpose particle detector
with a forward--backward symmetric cylindrical geometry and a near \(4\pi\) coverage in
solid angle.\footnote{\AtlasCoordFootnote}
It consists of an inner tracking detector surrounded by a thin superconducting solenoid
providing a \qty{2}{\tesla} axial magnetic field, electromagnetic and hadron calorimeters, and a muon spectrometer (MS).
The inner tracking detector (ID) covers the pseudorapidity range \(|\eta| < 2.5\).
It consists of silicon pixel, silicon microstrip, and transition radiation tracking detectors.
Lead/liquid-argon (LAr) sampling calorimeters provide electromagnetic (EM) energy measurements
with high granularity.
A steel/scintillator-tile hadron calorimeter covers the central pseudorapidity range of (\(|\eta| < 1.7\)).
The endcap and forward regions are instrumented with LAr calorimeters
for both the EM and hadronic energy measurements up to \(|\eta| = 4.9\).
The MS surrounds the calorimeters and is based on
three large superconducting air-core toroidal magnets with eight coils each.
The field integral of the toroids ranges between \num{2.0} and \qty{6.0}{\tesla\metre}
across most of the detector.
The MS includes a system of precision tracking chambers and fast detectors for triggering.
A two-level trigger system is used to select events.
The first-level trigger is implemented in hardware and uses a subset of the detector information
to accept events at a rate below \qty{100}{\kHz}.
This is followed by a software-based trigger that
reduces the accepted event rate to \qty{3}{\kHz} on average,
depending on the data-taking conditions.
An extensive software suite~\cite{ATL-SOFT-PUB-2021-001} is used in data simulation, in the reconstruction
and analysis of real and simulated data, in detector operations, and in the trigger and data acquisition
systems of the experiment.


\section{Data and simulated event samples}
\label{sec:samples}

The data and simulated event samples are summarised in the following.
 
\subsection{Data}
\label{sec:data}
 
The measurement presented in this paper is based on \pp collision data recorded by the ATLAS detector during the year 2022.
Data are selected only if all the detector components relevant for each channel were known to be in good operating condition.
The data samples correspond to an integrated luminosity of $31.4$ fb$^{-1}$ for the \hgg channel and of $29.0$ fb$^{-1}$ for \hfourl\ channel.
The difference between the sizes of the two data samples is due to additional requirements to remove data-taking periods with lower muon trigger efficiency applied only to the \hfourl\ channel.
The uncertainty in the integrated luminosity for data recorded in 2022 is 2.2\%~\cite{ATL-DAPR-PUB-2023-001}, following the methodology discussed
in Ref.~\cite{DAPR-2021-01}, using the LUCID-2 detector~\cite{LUCID2} for the primary luminosity measurements, complemented by measurements using the
ID and calorimeters.
The mean number of interactions per bunch crossing, called pile-up, averaged over all colliding bunch pairs is about 40.

Only events that satisfied single- or di-photon triggers~\cite{TRIG-2018-05} are used in the \hyy measurement. The single-photon trigger required
the presence of a photon candidate with at least 140~\GeV\ of transverse energy (\ET) in the event. This photon candidate is also required to
satisfy \textit{loose}~\cite{TRIG-2018-05} identification selections, as determined by properties of shower shapes in the calorimeters.
The di-photon trigger required
the presence of two photon candidates with \ET of at least 25 and 35~\GeV, and satisfying analogous \textit{medium}~\cite{TRIG-2018-05} identification requirements.
On average, these trigger requirements are 99.4\% efficient for the events that satisfy the analysis selection requirements
outlined in Section~\ref{sec:yy_reconstruction}. The efficiency is predominantly determined by the di-photon
trigger; the single-photon trigger adds less than 0.1\% of events of interest.
 
A total of 12 different single-lepton, dilepton, and trilepton triggers~\cite{TRIG-2018-01,TRIG-2018-05} were employed for the \hfourl measurement.
The single-lepton trigger thresholds used in 2022 were 24 and 26~\GeV\ for the muon and electron signatures, respectively.
The ensemble of the lepton triggers ensures a signal selection efficiency of 98.2\% for the events that satisfy the analysis selection requirements
outlined in Section~\ref{sec:zz_reconstruction}.

\subsection{Simulated event samples}
\label{sec:mcsamples}
 
At the LHC, Higgs bosons are primarily produced through gluon--gluon fusion (\ggF), vector-boson fusion (\VBF, where $V = W,Z$), and associated production with a
vector boson (\VH) and top- or bottom-quark pairs (\ttH, \bbH). These signal processes were simulated using
the \POWHEGBOX[v2]~\cite{Hamilton:2013fea,Hamilton:2015nsa,Alioli:2010xd,Nason:2004rx,Frixione:2007vw}
Monte Carlo (MC) generator.
Next-to-leading-order (NLO) accuracy was achieved for the \ggF and \VH processes with up to one extra parton by following
the \MINLO~\cite{Hamilton:2012np,Campbell:2012am,Hamilton:2012rf} approach,
while the $gg \to ZH$ process was only computed at leading order (LO) in quantum chromodynamics (QCD). The \VBF and \ttH processes were simulated at NLO accuracy in QCD\@.
The \bbH process was not simulated. Instead, since the acceptance of \ggF events is similar to that of \bbH events,
the \ggF cross-section is scaled to account for this contribution. The \bbH contribution accounts for approximately
1\% of the total \ggF+ \bbH production.
 
The samples are normalised to the state-of-the-art cross-section predictions~\cite{deFlorian:2016spz} interpolated to
the new LHC centre-of-mass energy of 13.6 \TeV. The extrapolation is based on cross-section values available at various centre-of-mass energies
ranging from 6 to 15 \TeV\ in 0.5 \TeV\ steps~\cite{LHCHXStwiki}.
The \ggF\ sample is normalised to a next-to-next-to-next-to-leading-order (N$^3$LO) QCD
calculation with NLO electroweak (EW) corrections~\cite{Aglietti:2004nj,Actis:2008ug,Actis:2008ts,Anastasiou:2008tj,Pak:2009dg,Harlander:2009bw,Harlander:2009mq,Harlander:2009my,Actis:2008ug,Anastasiou:2015ema,Anastasiou:2016cez,Dulat:2018rbf,Bonetti:2018ukf} for the \ggF process
summed with a \bbH\ contribution.
The \bbH\ contribution is normalised to a calculation that combines the complete NLO contributions that are present in the 4-flavour scheme calculation,
including finite $b$-quark mass effects and top-loop induced top-to-bottom Yukawa-coupling interference contributions, with the resummation of
collinear logarithms of $m_b/m_H$ as present in the 5-flavour scheme calculation up to next-to-next-to-leading-order (NNLO)~\cite{Forte_2015,Forte_2016,Bonvini_2015,Bonvini_2016}.
No EW corrections are included for \bbH.
The VBF sample is normalised to an approximate NNLO QCD
cross-section with NLO EW corrections~\cite{Ciccolini:2007jr,Ciccolini:2007ec,Bolzoni:2010xr}.
The $VH$ samples are normalised to cross-sections calculated at NNLO in
QCD with NLO EW corrections for $q\bar{q}/qg \to VH$ and at NLO
and next-to-leading-logarithm accuracy in QCD for $gg \to
ZH$~\cite{Ciccolini:2003jy,Brein:2003wg,Brein:2011vx,Altenkamp:2012sx,Denner:2014cla,Brein:2012ne,Harlander:2014wda,Harlander:2018yio}.
The \ttH\ cross-section is taken from a calculation accurate to NLO in QCD with NLO
EW corrections~\cite{Beenakker:2002nc,Dawson:2003zu,Yu:2014cka,Frixione:2015zaa}.
Summing up all the production modes,\footnote{When computing the total cross-section, the small
contribution from \tH\ production ($0.1$~pb) is also included.} the total predicted SM Higgs boson cross-section at 13.6~\TeV\ is \TotalCOMBExp.
 
All Higgs boson signal samples used the \PDFforLHC[21] set~\cite{Ball:2022a} of parton distribution functions (PDFs) in all matrix element (ME) calculations.
The samples were interfaced to \PYTHIA[8.2]~\cite{Sjostrand:2014zea} (\PYTHIA[8.3]~\cite{Bierlich2022}) to simulate \hyy (\hfourl) decays and to
model the effects of parton showering (PS), hadronisation and the underlying event (UE), with parameter values
determined by the A14 set of tuned parameters (tune)~\cite{ATL-PHYS-PUB-2014-021}.
The SM $H \to ZZ^* \to 4\ell~ (\ell =e,\mu)$  branching-ratio prediction is taken from \progname{Prophecy4f}~\cite{Bredenstein:2006rh,Bredenstein:2006nk}
and includes the full NLO EW corrections, as well as interference effects that result in a branching ratio that is
10\% higher for same-flavour final states ($4\mu$ and $4e$) than for different-flavour states ($2e2\mu$ and $2\mu2e$).
The samples were generated with a Higgs boson mass of 125.0~\GeV\footnote{A 90~\MeV\ shift is applied to the signal parameterisation used in the fit to the data.} and a Higgs boson width of 4.07~\MeV.
The predicted SM branching ratio is $(2.27 \pm 0.07) \times 10^{-3}$ for the $H \to \gamma \gamma$ decay~\cite{deFlorian:2016spz} and $(1.25 \pm 0.03) \times 10^{-4}$
for the $H \to ZZ^* \to 4\ell~ (\ell =e,\mu)$ decay,
for a Higgs boson mass of 125.09 \GeV.
 
Background simulated event samples for the \hyy measurement, consisting of non-resonant prompt di-photon production (\yy)
within a di-photon mass range of $90 < \mgg < 175$~\GeV, were generated with
\MGNLO~\cite{Alwall:2014hca} at an accuracy of NLO for MEs with up to two partons.
The \NNPDF[3.0] NLO PDF set~\cite{Ball:2014uwa} was used.
As with the Higgs boson signal samples, PS, hadronisation and UE
were modelled in \PYTHIA[8.2] with the A14 tune.
The merging of ME and PS was performed following the FxFx merging scheme~\cite{Frederix:2012ps}.
 
For the \hfourl measurement, the largest background contribution is from the $(Z^{(\ast)}/\gamma^{\ast})(Z^{(\ast)}/\gamma^{\ast})$ continuum background, jointly
called non-resonant or continuum \zzstar. The \zzstar\ continuum background from quark--antiquark annihilation (\emph{$qq\to ZZ^*$}) was
modelled using \progname{Sherpa}~2.2.12~\cite{Gleisberg:2008ta,Gleisberg:2008fv,Cascioli:2011va},
which provides a ME calculation accurate to NLO in QCD for 0- and 1-jet
final states, and LO accuracy for 2- and 3-jet final states.
The merging with the \progname{Sherpa} PS~\cite{Schumann:2007mg} was performed using the
ME+PS@NLO prescription~\cite{Hoeche:2012yf}.
Theoretical uncertainties in the $m_{4\ell}$ shape are evaluated by using Run~2 samples (generated at a centre-of-mass energy of 13 \TeV) and comparing the baseline \progname{Sherpa} 2.2.2 sample with a
\progname{MadGraph5\_aMC@NLO} sample that uses \progname{FxFx} merging at NLO for 0- and 1-jet final states, interfaced to \progname{Pythia~8} for PS.
 
The gluon-induced \zzstar\ production is taken into account by scaling the expected \zzstar\ continuum background from quark--antiquark
annihilation to the expected cross-section for the sum of the two processes. This scaling is derived using Run~2 simulated event samples (generated at a centre-of-mass energy
of 13 \TeV) with \progname{Sherpa} 2.2.2 and also takes into account the slightly different mass shapes of the two processes.
Since the total normalisation of the \zzstar\ sample is derived from the fit to the data
(as described in the following), only the relative normalisation, used to determine the invariant mass shape, is relevant to this study.
The higher-order QCD effects for the $gg\rightarrow ZZ^{*}$ continuum production are calculated for massless quark loops~\cite{Caola:2015psa,Caola:2015rqy,Campbell:2016ivq} in the heavy top-quark
approximation~\cite{Melnikov:2015laa}, including the $gg\rightarrow H^{*} \rightarrow ZZ^{*}$ process~\cite{Bonvini:2013jha,Li:2015jva}.
The $gg \to ZZ^*$ LO cross-section was scaled by a $K$-factor of 1.7$\pm$1.0, defined as the ratio of the higher-order to leading-order cross-section predictions.
Production of \zzstar\ via vector-boson scattering was simulated at LO in QCD with the \progname{Sherpa} 2.2.2 generator.

The \emph{WZ} background was modelled using  the \progname{Sherpa} 2.2.12 generator using the same set-up as for the \emph{$qq\to ZZ^*$} sample. The triboson backgrounds \emph{ZZZ}, \emph{WZZ}, and \emph{WWZ} with four or more prompt leptons  were modelled using \progname{Sherpa} 2.2.12. Other sub-dominant backgrounds like the production of vector bosons in association with a top-quark pair ($t\bar{t}W/Z$) were simulated using the \progname{Sherpa} 2.2.12 generator.

Events containing $Z$ bosons with associated jets (\zjets)
were simulated using the \progname{Sherpa}~2.2.12 generator. MEs were calculated for up to two partons at NLO and four partons at LO using
\progname{Comix} \cite{Gleisberg:2008fv} and \progname{OpenLoops} \cite{Cascioli:2011va},
and merged with the \progname{Sherpa} PS~\cite{Schumann:2007mg} using
the \progname{ME+PS@NLO} prescription~\cite{Hoeche:2012yf}. The NNPDF3.0 NNLO PDF set was used in conjunction
with a dedicated set of tuned PS parameters.
 
The $t\bar{t}$ background was modelled using \POWHEGBOX[v2] interfaced to \PYTHIA[8.3]
for PS, hadronisation, and the UE, and to \progname{EvtGen} v2.1.1 for heavy-flavour hadron decays~\cite{Frixione:2007nw,Nason:2004rx, Frixione:2007vw,Alioli:2010xd}. For this sample,
the A14 tune~\cite{ATL-PHYS-PUB-2016-020} and NNPDF3.0 NLO PDFs were used.
Simulated \zjets\ and  $t\bar{t}$ background event samples are normalised to the data-driven estimates described in Section~\ref{sec:zz_bkg}.
 
All generated signal and background events were passed through a simulation of the ATLAS detector
using \GEANT~\cite{Agostinelli:2002hh,SOFT-2010-01}.
Scale factors are applied to the simulation to match the performance of the different physics objects to those measured in the data.
 
A summary of the simulated event samples used in this measurement together with the total cross-section values used for the normalisation
of signal processes in the full phase space is provided in Table~\ref{tab:samples}.
All the background processes reported in this table (except the small irreducible $VVV$ and $t\bar{t} W/Z$) are directly normalised to
the data or derived from data control regions.
 
The effect of multiple interactions from the same or neighbouring bunch crossings was modelled
through the overlay~\cite{SIMU-2020-01} of the hard-scatter signal and background events with a set of simulated inelastic
\pp events generated by EPOS 2.0.1.4~\cite{Porteboeuf:2010um}
and \PYTHIA[8.2]. Events with at least one high transverse-momentum (\pT) jet,
isolated photon, or lepton from the decay of a $b$-hadron were simulated using \PYTHIA[8.2]. All other events were simulated using EPOS;
a veto on EPOS events with such a high transverse-momentum signature ensured orthogonality with the \PYTHIA[8.2] sample.
The EPOS LHC tune was used to set the parameters of the EPOS generation,
while the \PYTHIA[8.2] parameters were set using the A3 tune~\cite{ATL-PHYS-PUB-2016-017}. The
\NNPDF[2.3 LO] set of PDFs~\cite{Ball:2012cx} was used.
Dedicated weights were used to ensure that the mean number of interactions per bunch crossing observed in the 2022 data
is adequately described in these samples.
 
\begin{table}[!htbp]
\centering
\caption{Simulated signal and background event samples. The last column shows the total cross-sections used to normalise the signal
samples for $m_{H} = 125.09$ \GeV. The uncertainties on the cross-sections are based on the theoretical papers cited in the
text and include several components as described in Ref.~\cite{deFlorian:2016spz} (missing higher order QCD and EW corrections, PDF, strong coupling
constant, quark mass effects, etc.).}
\label{tab:samples}
\renewcommand{\arraystretch}{1.3}
\resizebox{\textwidth}{!}{
\begin{tabular}{lccccrrr}
\toprule
\multirow{2}{*}{Process} & \multirow{2}{*}{Generator} & \multirow{2}{*}{Showering} & \multirow{2}{*}{PDF set} & $\sigma$ [pb]  \\
& & & & $\sqrt{s}$ = 13.6~\TeV \\
\midrule
\ggF + \bbH &  \POWHEGBOX[v2] + \MINLO & \PYTHIA[8.2/8.3] & PDF4LHC21 &  $52.7 \pm 2.6$ \\
\VBF & \POWHEGBOX[v2] & \PYTHIA[8.2/8.3] & PDF4LHC21 &  $4.075 ^{+0.088}_{-0.089}$ \\
\WH & \POWHEGBOX[v2] + \MINLO & \PYTHIA[8.2/8.3] & PDF4LHC21 & $1.453^{+0.029}_{-0.028}$ \\
$q\bar{q}\to ZH$ & \POWHEGBOX[v2] + \MINLO & \PYTHIA[8.2/8.3] & PDF4LHC21 & $0.806^{+0.033}_{-0.029}$ \\
$gg\to ZH$ & \POWHEGBOX[v2] & \PYTHIA[8.2/8.3] & PDF4LHC21 & $0.136^{+0.034}_{-0.026} $ \\
\ttH & \POWHEGBOX[v2] & \PYTHIA[8.2/8.3] & PDF4LHC21 &  $0.569^{+0.040}_{-0.057}$  \\
\hline
\yy, $m_{\gamma\gamma} \in$ 90--175~\GeV & \MGNLO & \PYTHIA[8.2] & \NNPDF[3.0] NLO & - \\
$qq\to$\zzstar\  & \progname{Sherpa} 2.2.12  & \progname{Sherpa} 2.2.12 & \NNPDF[3.0] NNLO & - \\
$VV$, $VVV$ & \progname{Sherpa} 2.2.12  & \progname{Sherpa} 2.2.12 & \NNPDF[3.0] NNLO & - \\
\zjets &  \progname{Sherpa} 2.2.12  & \progname{Sherpa} 2.2.12 & \NNPDF[3.0] NLO& - \\
$t\bar{t}$ &  \POWHEGBOX[v2] & \PYTHIA[8.3]  & \NNPDF[3.0] NLO & - \\
$t\bar{t}W$, $t\bar{t}Z$ &  \progname{Sherpa} 2.2.12 &  \progname{Sherpa} 2.2.12 &  \NNPDF[3.0] NNLO  & - \\
\bottomrule
\end{tabular}
}
\end{table}


\section{The \hyy measurement}
\label{sec:hyy}
 
\subsection{Event reconstruction and selection}
\label{sec:yy_reconstruction}

The events used in the \hyy measurement are selected using a multi-step procedure.
Reconstructed photon candidates are first required to satisfy a set of preselection requirements.
Subsequently, they are ranked according to their transverse energy and the two
highest-\ET candidates are used to build a di-photon
system and determine the primary vertex of the event.
A final set of requirements is then applied to the di-photon system.
 
\subsubsection{Photon reconstruction, identification and preselection}
\label{sec:photons}
 
Photon reconstruction is performed using a dynamic clustering algorithm~\cite{ATL-PHYS-PUB-2017-022,EGAM-2018-01}
in which variable-sized topological clusters~\cite{PERF-2014-07} are built from
signals significantly above the noise level in EM-calorimeter cells. Additional clusters
consistent with energy deposits from the products of photon conversion are added
to the original cluster if relevant.
Tracks reconstructed in the ID
are evaluated based on the quality of their matching to the reconstructed calorimeter-energy cluster.
Photon candidates with no matching track are labelled unconverted photons while photon
candidates with a matching track with no innermost pixel-layer hits or with two matching
tracks are classified as converted photons.
All photon candidates are required to be within the acceptance region of the finely segmented
first layer of the EM calorimeter,
$|\eta| < 2.37$, and outside of the area that falls between the barrel and endcap calorimeters,
$1.37 < |\eta| < 1.52$.
Their energy measurement is calibrated using the method described in Ref.~\cite{EGAM-2018-01}.
 
The procedure described above is effective at reconstructing not only prompt photons but also
other similar objects, such as di-photon decays of neutral hadrons inside jets.
These other objects are rejected using an identification algorithm based
on calorimeter shower-shape information~\cite{PERF-2013-04}. The \textit{loose}~\cite{EGAM-2018-01} working point
serves as the initial preselection requirement.
It is based on the shape of the shower in the second layer of the EM calorimeter
and on the amount of energy deposited in the hadronic calorimeter, and has a nominal prompt-photon efficiency of above 98\%
for prompt photons with transverse momentum greater than 30~\GeV.
 
An additional preselection requirement retains events with at least two photon candidates,
each with $\ET > 25$~\GeV, and a di-photon system is built. 
In events where more than two photon candidates satisfy these preselection requirements, the
two highest-\ET candidates are used to construct the di-photon system.\footnote{The two highest-\ET
candidates match the Higgs boson decay products in over 99\% of simulated signal events.}
 
\subsubsection{Primary vertex selection}
\label{sec:yy_vertexselection}
 
The primary vertex of the event is selected using a neural-network algorithm~\cite{HIGG-2013-08} trained to
distinguish between the hard-scatter vertex of \ggF events and vertices
from pile-up interactions.
The algorithm makes use of information
about all reconstructed vertices in the event\footnote{Reconstructed vertices are required to have at least two associated ID
tracks of transverse momentum $\pt > 500$~\MeV.} and about
the kinematic properties of the two photons forming the di-photon system,
as measured in the calorimeter and complemented, in the case of
converted photons, by kinematics of matching ID tracks.
The use of this neural network provides a vertex selection efficiency of 71.4\% for the most abundant ggF process,
to be compared with an efficiency of 52.2\% when the primary vertex is selected solely based on the sum of the \pT-squared
of its associated tracks, which is generally the default in ATLAS\@.
This efficiency is estimated by using MC simulation, requiring a distance along the $z$-axis between the particle-level primary vertex
and the reconstructed one of less than 0.3~mm.
Agreement between data and simulation is validated in the Run 3 data sample using $Z\to ee$ events, treating the electrons as
unconverted photon candidates (i.e., ignoring their track information). A weak dependence
of the performance on pile-up levels is observed, but it is well-modelled
by the simulation.
 
Once the primary vertex is selected, the direction of the two
photon candidates is re-computed. This has the benefit of improving the
\ET measurement of the photon candidates through its dependence on the photon candidate's
direction in $\eta$. It also improves the di-photon invariant mass resolution by about 8\%.
 
\subsubsection{Event selection}
The event selection is completed by a set of requirements on the two photon candidates
that constitute the di-photon system.
 
First, a matching requirement is applied between the selected photon candidates and the corresponding photons at trigger level,
as described in Section ~\ref{sec:data}.
Either one of the photon candidates must be matched to the single-photon trigger, or both photon candidates
must be matched to the di-photon trigger.
The leading \ET and sub-leading \ET photon candidates are additionally required to have $E_T / \mgg > 0.35$
and $>0.25$ respectively.
 
At this stage, a more stringent \textit{tight}~\cite{PERF-2017-02} working point requirement is applied.
The efficiency of the \textit{tight} working point requirement
varies with the \ET of the photon, ranging from 84\% (85\%) at $\ET = 25$~\GeV\ to
94\% (98\%) at $\ET = 100$~\GeV\ for unconverted (converted) photons.
This working point makes use of the same information as the \textit{loose} working point used for the
preselection, albeit with more stringent requirements, and adds information from the finely segmented
first layer of the EM calorimeter.
 
To further reject hadronic background, surviving photon candidates are required to be
isolated from other significant activity in the ID or the calorimeters.
Track-based isolation is determined by the scalar sum of the \pT of all tracks of
at least 1~\GeV\ that are within a cone of $\Delta R = 0.2$ of the photon candidate and are
compatible with originating from the selected primary vertex.
Track isolation requirements are satisfied if this \pT-sum is less than 5\% of the
photon candidate's \ET.
Calorimeter-based isolation, on the other hand, is defined as the scalar sum of the
\ET of all positive-energy topological clusters within the same $\Delta R = 0.2$ cone, excluding
the energy of the photon candidate itself. For calorimeter-based isolation, an additional
subtraction is performed to remove pile-up and UE contributions. This uses an
ambient energy correction based on the low-\pT jets of the event~\cite{ATL-PHYS-PUB-2017-022,Cacciari:2007fd}.
Calorimeter isolation requirements are satisfied if the resulting \ET-sum is
less than 6.5\% of the photon candidate's \ET.
Both the track and calorimeter isolation requirements must be satisfied.
 
The last requirement restricts the invariant mass \mgg of the two
photon candidates to be within the range 105--160~\GeV.
 
The number of selected events in the 2022 Run 3 data sample is 307 996.
The selection efficiency, relative to the full phase space, is estimated
from simulated events to be 36\%. 
 
\subsubsection{Fiducial region}
\label{sec:yy_fiducial}
The fiducial region considered in this measurement is designed to closely match the
selection requirements outlined above to minimise model dependence
due to acceptance extrapolations. The di-photon fiducial region is defined by the
presence of two prompt photons with transverse momentum greater than 0.35 and 0.25 times
the di-photon invariant mass. The photons are required to be in the range of $|\eta| < 2.37$
but outside of the region $1.37 < |\eta| < 1.52$.
They are also required to be isolated, based on a particle-level isolation criterion defined as the transverse energy obtained
from the vectorial sum of the momenta of all the stable charged
particles (mean lifetime $c\tau > 10$ mm) with \pT
greater than 1~\GeV\ and within a cone of $\Delta R = 0.2$ around the photon's direction.
This quantity must be less than 5\% of the photon's \ET.
The invariant mass of the di-photon system is required to be
within the range 105--160~\GeV. These requirements are summarised in Table~\ref{tab:yy_fid_def}.
 
The signal acceptance of this di-photon fiducial region is 50\%
relative to the full phase space. In total, 2117 signal events are expected in the fiducial region assuming
SM values for the production cross-section, the acceptance and the di-photon branching fraction.
About 1.9\% of the events that
satisfy the detector-level event selection fail to satisfy the fiducial selection.
 
\begin{table}[!htbp]
\centering
\caption{Summary of the event selection requirements that define the fiducial phase space for the \hyy cross-section measurement.}
\label{tab:yy_fid_def}
\begin{tabular}{ll}
\toprule
\multicolumn{2}{c}{\textbf{Photons}} \\
\midrule
Leading (sub-leading) $p_\mathrm{T}^{\gamma}$ & $ p_\mathrm{T}^{\gamma} / \mgg > 0.35 (0.25) $ \\
Pseudorapidity & $|\eta| < 2.37$ and outside $1.37 < |\eta| < 1.52$ \\
Isolation ($\Delta R = 0.2 $)& $E_\mathrm{T}^\mathrm{iso}/E_\mathrm{T}^\gamma < 0.05$ \\
\midrule
\multicolumn{2}{c}{\textbf{Di-photon system}} \\
\midrule
Mass window & $105 \text{\GeV} < m_{\gamma\gamma} < 160$~\GeV \\
\bottomrule
\end{tabular}
\end{table}


\subsection{Fit procedure}
\label{sec:yy_modeling}

The methodology used to extract the $pp \to H \to \gamma\gamma$ fiducial cross-section
closely follows that of a previous ATLAS measurement~\cite{HIGG-2019-13}.
The cross-section is measured via an analytic fit to the di-photon mass spectrum.
The analytic function used to model the signal distribution is derived mainly from simulation, after applying the relevant calibrations, while the
parameters of the analytic function used to model the background distribution are determined from a fit to the data.
For this inclusive measurement, the measured signal yields are corrected to particle level using a single
correction factor derived from MC simulation. This correction factor takes into account detector effects and
out-of-acceptance signal contributions.

The fiducial cross-section is extracted via an unbinned maximum-likelihood fit to the reconstructed di-photon invariant mass spectrum in
the range 105--160~\GeV.
Confidence intervals are based on the profile-likelihood-ratio test statistic~\cite{Cowan:2010st}.
The systematic uncertainties listed in Section~\ref{sec:yy_systematics} are implemented in the fit as nuisance
parameters constrained by additional Gaussian or log-normal likelihood terms.
The probability density functions for the signal and background are described by analytic functions
detailed in Sections~\ref{sec:yy_smodel} and \ref{sec:yy_bmodel}.
In the likelihood, the signal yield $N_S$ is parameterised in terms of the fiducial cross-section as
\begin{equation}
\label{eq:NS}
N_{S} = \sigma_\mathrm{fid} \times \cal{L} \times \cal{C_{F}},
\end{equation}
where $\sigma_\mathrm{fid}$ is the fiducial cross-section, the parameter of interest of the fit, and is defined as
the product of the total production cross-section $\sigma_\mathrm{tot}$, the di-photon branching ratio $\cal{B}_{\gamma \gamma}$ and the acceptance $\cal{A}$:
\begin{equation}
\label{eq:XS}
\sigma_\mathrm{fid} = \sigma_\mathrm{tot} \times \cal{B}_{\gamma \gamma} \times \cal{A}.
\end{equation}
The signal acceptance $\cal{A}$ is the fraction of signal events that passes the particle-level fiducial selection described in
Section~\ref{sec:yy_reconstruction}.
The other terms of Eq.~(\ref{eq:NS}) are $\cal{L}$, the integrated luminosity, and $\cal{C_{F}}$,
the correction factor, defined as the ratio of the number of signal events selected at detector level
to the number of signal events selected at particle level after applying
the selection criteria described in Section~\ref{sec:yy_reconstruction}.
The correction factor is computed inclusively assuming that the ratio of the different
Higgs boson production modes is as in the SM\@. The value of $\cal{C_{F}}$, derived from the signal simulated event samples, is 71.6\%.
The dependence on the production mode is very weak and ranges from 71.5\% for \ggF to 72.4\% for \VBF\@.\footnote{A larger value of 75\% for the correction factor is obtained for the $gg \rightarrow ZH$ process, but in the SM this process only accounts for 0.3\% of the selected signal yield.}

\subsection{Signal modelling}
\label{sec:yy_smodel}
The signal mass shape is modelled with a double-sided Crystal Ball function, detailed in Ref.~\cite{HIGG-2016-21}.
This function is characterised by a Gaussian component that describes the central part of the mass spectrum and two independent
power-law tails. In total the signal function is described by six shape parameters and one overall-normalisation parameter.
Among the shape parameters are the position of the signal peak, linked to the Higgs boson mass, and the experimental resolution.
They can be mapped to the mean $\mu_\mathrm{CB}$
and the width $\sigma_\mathrm{CB}$ of the central Gaussian component.
Another four parameters determine the shape of the power-law tails.
These six Crystal Ball shape parameters are determined from a fit to the signal simulated event samples
and are kept fixed in the fit to the data within the associated systematic uncertainties described in Section~\ref{subsec:masssyst}.
Lastly, the overall normalisation is expressed by the parameter $N_{S}$ previously defined.
It is determined in the fit to the data.
The validity of the signal parameterisation is checked by a `closure test' performed with a fit to a simulated data sample
obtained by injecting known signal yields on top of a background template.
 
Since the signal samples are generated at a Higgs boson mass of 125.0 \GeV, the mean value of the central Gaussian
component used in the signal parameterisation is shifted by 90 \MeV\ to emulate a signal at 125.09 \GeV~\cite{HIGG-2014-14}.
 
Contributions from the different Higgs boson production modes are added together according to the yield fractions predicted by the SM\@.
No significant differences are found between the signal shapes of the different production modes,
indicating that the signal parameterisation is not sensitive to assumptions on their relative normalisation.
 
Several sources of systematic uncertainty associated with the signal modelling are taken into account. They are described
in Section~\ref{sec:yy_systematics}.

\subsection{Background modelling}
\label{sec:yy_bmodel}
Analytic models are also used to describe the various background contributions.
The functional form chosen for the background model is assessed using background templates built using
MC simulation and data control regions. These templates are also used for the estimation of related systematic uncertainties.
 
\subsubsection{Background templates}
\label{subsec:yy_bkgtemplate}
The main sources of background are the non-resonant production of prompt and isolated di-photons (\gamgam)
and the \gamjet\footnote{The \gamjet\ estimate includes both cases where either the leading \ET or the sub-leading \ET photon are from misidentified jets.}  and \jetjet\ processes where one or two hadronic jets $j$ are misidentified as photons.
The relative contribution of each background component is measured in data using
a two-dimensional double-sideband method~\cite{STDM-2011-05}.
With this method the fraction of the three components is measured to be $73.0\pm 1.9$\% for the \gamgam, $23.6 \pm 1.5$\%
for the \gamjet, and $3.4 \pm 0.4$\% for the \jetjet process.
 
A background template of the di-photon invariant mass is built by summing the \gamgam, \gamjet\
and \jetjet\ components according to the measured fractions.
The \gamgam\ component is directly derived from the simulated background event sample
after applying the event selection described in Section~\ref{sec:yy_reconstruction}.
The shape of the sum of the \gamjet\ and \jetjet\ components is derived from data control samples defined by
inverting the identification and/or isolation requirements of one or both photons. Since the ratio of the \mgg\ distribution
of the \gamjet\ and \jetjet\ components to that of the simulated \gamgam\ event
sample is found to be described by a simple linear function in \mgg, a linear fit to the ratio of these distributions
is used to provide an \mgg-dependent weight that is
applied to the \gamgam\ sample to obtain a template that accounts for both the \gamjet\ and \jetjet\ components.

To get a better description of the background, a second-order polynomial in \mgg\ is fitted to
the ratio of the template to the data in the sidebands (i.e., excluding the 120--130~\GeV\ mass region) and is used to
correct the background template shape.
 
A second, alternative template used for background modelling studies is created through the implementation of
a normalising flow (NF) method~\cite{Xu:2023xdc} that uses a generative machine-learning model based on MADE~\cite{NF_MADE} and MAF~\cite{NF_Mask},
developed to emulate the detector response to photons and applied to
particle-level simulated \gamgam\ events.
This method is much faster than the standard full \GEANT\ simulation,
and can be applied to very large samples. The alternative template is produced using the equivalent of about
3.5~ab$^{-1}$ of integrated luminosity, compared to the 1.3~ab$^{-1}$ of the \GEANT -based sample.
To correct for the photon identification and isolation
efficiencies, which are not part of the learning procedure, a parameterised photon
efficiency derived as a function of the photon transverse energy and pseudorapidity from the full simulation is applied to
the NF simulated events.
The prompt di-photon background component obtained with the NF approach is then corrected to take into account the non-prompt
photon contributions using a procedure identical to the one described above for the \GEANT-based sample,
including the reweighting of the simulated mass spectrum with a second-order polynomial to the data sidebands.
 
\subsubsection{Functional form}
The \mgg\ background distribution is characterised by a smoothly falling spectrum. It can be parameterised by an empirical function
selected using the background templates described above. The parameters of this function
are fitted to data, but the functional form itself is chosen following a procedure, used in a previous publication~\cite{HIGG-2019-13}, which aims
to minimise the statistical and systematic uncertainties in the measured signal while still
ensuring a good description of the background \mgg spectrum.
 
Three families of functions are studied for this empirical component: power-law functions, Bernstein polynomials and
exponential functions of polynomials.
The criteria used to select the background function from among these families and to determine the associated systematic uncertainty,
referred to as the spurious signal (SS)~\cite{HIGG-2012-27}, are described in the following. The procedure begins by
considering functions of lowest order; the order is increased until a function satisfying the criteria is found.
 
As a first requirement, only functions able to fit the data mass sidebands with a $\chi^{2}$ probability
greater than $1\%$ are considered. 
For all functions that satisfy this criterion, a spurious-signal test is performed.
The SS yields are defined as the measured signal obtained in signal-plus-background fits
to the background-only templates. They are determined not only for a signal
mass of $125\,\GeV$, but for all values of $m_H$ between 123 and 127\,\GeV\ in
steps of $0.5\,\GeV$. The maximum absolute value of the SS yield in this mass
range is taken to be the background modelling uncertainty.
Only functions with an absolute value of the SS yield below 20\% of the statistical error of the expected fitted signal yield
or below 10\% of the expected SM signal yield are considered.
Among the functions that satisfy this requirement, those with the smallest number of
degrees of freedom are selected to minimise the statistical uncertainty that dominates this
measurement. If more than one function passes this final selection, the one with
the smallest SS is selected.
As a conservative approach, the SS is computed using the two background templates (\GEANT- and NF-based)
described in Section~\ref{subsec:yy_bkgtemplate} and the largest value is selected.
The function selected by this procedure is an exponential function of a second-order polynomial in \mgg,
with an associated SS uncertainty corresponding to 6.2\% of the expected SM signal yield.
 
To test the robustness of the background modelling choice, the procedure described above is repeated on
different background templates.
These additional templates are obtained using only the prompt-photon simulation reweighted to the data sidebands,
either from the full-simulation samples or from the particle-level samples corrected for detector effects with the
NF approach.
For these additional templates, the procedure still selects the same function and the values of the SS
are smaller than those of the original templates described in Section~\ref{subsec:yy_bkgtemplate}, hence no additional uncertainty is associated with this test.


\subsection{Systematic uncertainties}
\label{sec:yy_systematics}

The measurement of the fiducial \hyy cross-section is affected by several sources of uncertainty,
which can be broken down into three categories:
those that affect the shape of the \mgg\ signal distribution,
those that affect the background modelling,
and those that impact the detector and fiducial corrections to the measured signal yields and thus the correction factor $\cal{C_{F}}$.
The uncertainty in the luminosity measurement also
affects the fiducial cross-section measurement, as illustrated by the relation in Eq.~(\ref{eq:NS}).
This uncertainty, described in Section~\ref{sec:data},
is evaluated to be 2.2\%.
 
\subsubsection{Systematic uncertainties in the \mgg signal modelling}
\label{subsec:masssyst}
Since the fiducial cross-section is measured from a fit to the \mgg spectrum, any sources of uncertainty that affect the peak
position or the width of the signal peak will have an impact on the cross-section measurement. Uncertainties in the photon-energy scale
result in a shift of the position of the signal peak, while uncertainties in the photon-energy resolution result in the broadening
or narrowing of this peak.
These uncertainties are evaluated as described in Refs.~\cite{EGAM-2018-01,PERF-2017-03} using Run~2 data and simulation. 
Additional uncertainties are included to
take into account the change in conditions between Run~2 and Run~3, and changes to the MC simulation.
The uncertainty arising from differences in the photon-energy scale between data and simulation has a modest impact of under 0.25\% on the position of the peak;
an additional 0.4\% uncertainty accounts for changes in the detector simulation between Run~2 and Run~3.
The photon-energy resolution uncertainty is the single largest contributor to the signal shape uncertainty, with
a total error of about 10\% in the width of the peak.
 
The peak position is also affected by the uncertainty in the Higgs boson mass, taken as 0.24~\GeV\ from the combination of the
Higgs boson mass measurements performed by the ATLAS and CMS collaborations during the first run of the LHC~\cite{HIGG-2014-14}.
This uncertainty has an impact on the cross-section measurement of 0.1\%.

\subsubsection{Systematic uncertainties in the \mgg background modelling}
The choice of the analytic function used to model the background \mgg distribution can affect the measured
cross-section, and the associated impact is quantified through the spurious signal yield described in Section~\ref{sec:yy_bmodel}.

\subsubsection{Systematic uncertainties in the $\cal{C_{F}}$ correction factor}
 
Both experimental and theoretical uncertainties can affect the correction factor $\cal{C_{F}}$,
and in the case of theoretical uncertainties, introduce a slight model dependence into the measurement.
These sources of uncertainty also have an impact on the fiducial acceptance, and they are relevant
when extrapolating the fiducial cross-section measurement results to the
full phase space.
 
Several experimental uncertainties impact the $\cal{C_{F}}$ correction factor.
These include uncertainties related to the efficiency of the photon trigger,
identification and isolation selections, to the impact of the di-photon vertex selection, and to the estimation of the photon-energy scale and resolution.
 
The efficiency of the di-photon trigger is estimated with an uncertainty of
about 2\% for Run~3 data based on the bootstrap method~\cite{TRIG-2018-05,TRIG-2016-01}.
The single photon trigger contributes only to 0.1\% of the
selected signal events and its uncertainty is therefore neglected.
All other photon-related uncertainties are assessed
using Run~2 data and additional uncertainties take into account effects from the change
in conditions between Run~2 and Run~3, and changes to the MC simulation.
 
The photon identification efficiency in simulated events is corrected to match the efficiency observed in data as detailed
in Ref.~\cite{EGAM-2018-01}. The uncertainty in the simulation-to-data
correction is estimated to have an impact of about 5\% on $\cal{C_{F}}$.
The uncertainty in the photon isolation efficiency is derived from a similar methodology to that of the photon identification
as detailed in Ref.~\cite{EGAM-2018-01}.
The photon isolation efficiency uncertainty is estimated to have a 1.6\% impact on $\cal{C_{F}}$.
The efficiency of the vertex selection algorithm is computed both in data
and in simulation. As described in Section~\ref{sec:yy_reconstruction}, the difference between the two is evaluated using
$Z\to ee$ events. The resulting impact on $\cal{C_{F}}$ from this efficiency uncertainty amounts to below 0.03\%.
 
Changes to the photon-energy scale and resolution result in the migration of
events into and out of the fiducial region, with a maximal impact on
$\cal{C_{F}}$ below 0.1\%.
 
The uncertainty in the modelling of pile-up is evaluated by varying by $\pm 9\%$ the value of the visible inelastic
cross-section used to reweight the pile-up distribution in the simulation to that observed in the data~\cite{STDM-2015-05}.
The uncertainty in the modelling of pile-up is estimated to have an impact of about 1.2\% on $\cal{C_{F}}$.

Several sources of theoretical uncertainty are also considered. The correction factor $\cal{C_{F}}$ is computed by combining all production modes of the Higgs boson
assuming SM cross-sections. Values of $\cal{C_{F}}$ that are different between production
modes introduce a model dependence through the SM cross-section assumption. A signal composition
uncertainty is therefore derived by varying the cross-section of each production mode within its measured
uncertainty as determined in Ref.~\cite{HIGG-2021-23}. The impact of this uncertainty on
$\cal{C_{F}}$ is 0.02\%.
Uncertainties in $\cal{C_{F}}$ due to the ME generator, scale and PDF set choice,
and the modelling of the PS, UE and hadronisation, are all small and
neglected in the fit model.
 
Additional uncertainties are considered when extrapolating the fiducial cross-section to the full phase space to derive the
total cross-section. They include uncertainties in the acceptance that are derived following the same procedure as in
Refs.~\cite{HIGG-2020-16,HIGG-2019-13}, mainly originating from missing higher-order terms in the perturbative QCD calculations,
the modelling of PS, the PDFs and the value of \alphas, for a total amount of 1.4\%.
An uncertainty in the \hyy\ branching ratio of 2.9\%~\cite{deFlorian:2016spz} is also included.


\subsection{Results}
\label{sec:yy_results}

The invariant mass spectrum, superimposed with the background and signal functions obtained from the fit to the data,
is shown in Figure~\ref{fig:massfit}.
 
\begin{figure}[htbp]
\centering
\includegraphics[width=0.8\textwidth]{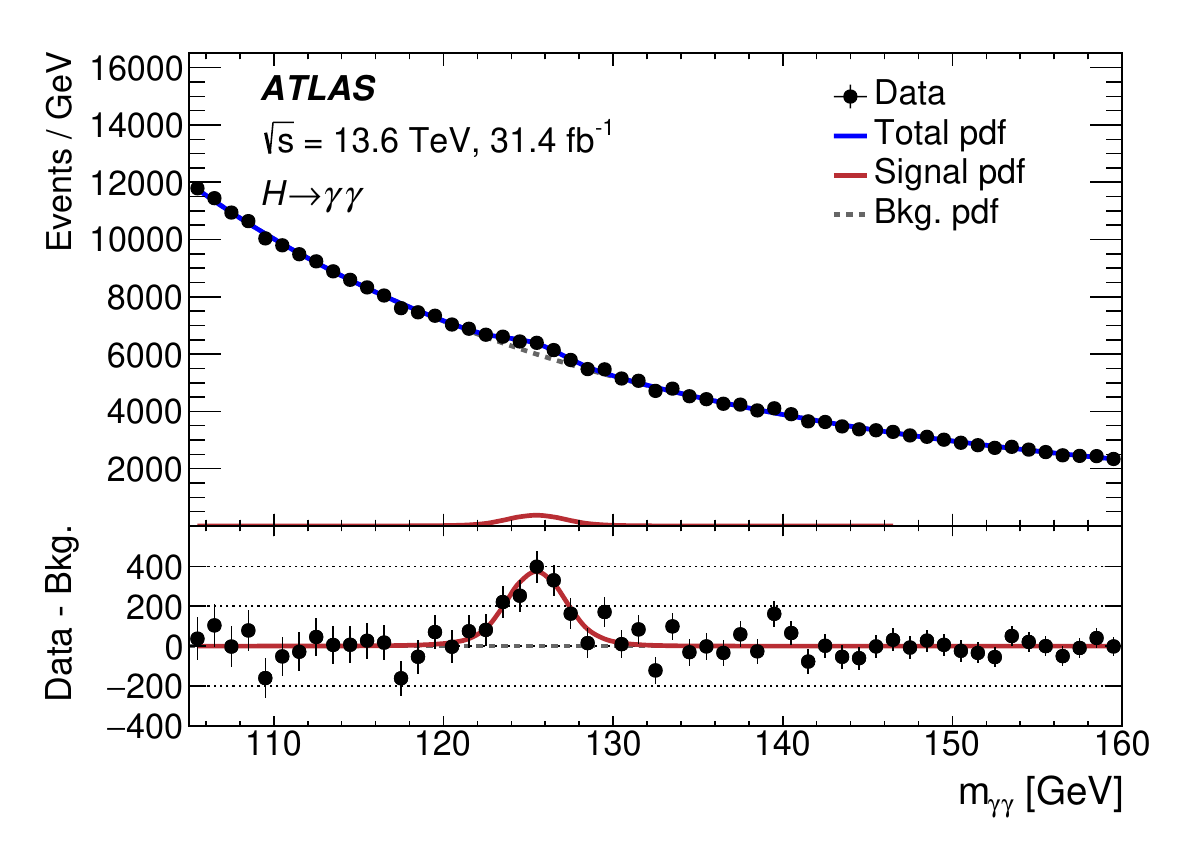}
\caption{ Di-photon invariant mass spectrum within the \hgg fiducial region, as observed in data. The signal, background and total probability density functions (pdf)
derived from the fit to data in the window $105 \le  \mgg \le 160$~\GeV\ are also shown.
}
\label{fig:massfit}
\end{figure}
 
The value of the fiducial cross-section, extracted from the fit to this \mgg\ spectrum, is measured to be \FiducialYY\ and is
in agreement with the SM prediction of \FiducialYYExp. The SM prediction is derived using the cross-section
calculations described in Section~\ref{sec:mcsamples} and corrected for acceptance effects, assessed using the simulated event samples.
 
The breakdown of the total uncertainty into its different components is detailed in Table~\ref{tab:syst_summary_diff}.
The statistical component is obtained by fixing all the nuisance parameters associated with systematic uncertainties to their best-fit values.
The grouped squared systematic uncertainties are obtained by subtracting the squared values of the uncertainties obtained from two fits,
one performed with the corresponding nuisance parameters fixed to their best-fit value and the other with them free floating in the fit.

\begin{table}[htbp]
\centering
\caption{Breakdown of the relative uncertainties in the inclusive di-photon fiducial cross-section measurement.}
\begin{tabular}{lS[table-format=< 2.1,table-align-comparator = false]}
\toprule
Source  &  {Uncertainty [\%]} \\
\hline
Statistical uncertainty  & 14.0 \\
Systematic uncertainty   & 10.3 \\
\quad Background modelling (spurious signal) & 6.0 \\
\quad Photon trigger and selection efficiency & 5.8 \\
\quad Photon energy scale \& resolution & 5.5 \\
\quad Luminosity & 2.2 \\
\quad Pile-up modelling & 1.2 \\
\quad Higgs boson mass  & 0.1 \\
\quad Theoretical (signal) modelling & < 0.1 \\
\hline
Total & 17.4\\
\bottomrule
\end{tabular}
\label{tab:syst_summary_diff}
\end{table}


\section{The \hzz measurement}
\label{sec:h4l}
 
\subsection{Event reconstruction and selection}
\label{sec:zz_reconstruction}

The selection of events for the \hfourl measurement relies on the reconstruction and
identification of lepton (electron and muon) candidates, used to build a quadruplet
that defines the Higgs boson candidate.
It closely follows the analyses reported in Refs.~\cite{HIGG-2018-28, HIGG-2018-29}.
A brief description is provided below.
 
\subsubsection{Electron and muon reconstruction, identification and preselection}
\label{sec:zz_emureco}
 
Electron reconstruction is performed using energy clusters in the EM calorimeter built with a dynamical,
topological cell-based approach~\cite{ATL-PHYS-PUB-2017-022,EGAM-2018-01} similar to the one described for photons in
Section~\ref{sec:photons}. These clusters are required to be matched to reconstructed ID tracks.
A Gaussian-sum filter algorithm~\cite{ATLAS-CONF-2012-047} is used to compensate for radiative energy losses in the ID for the track reconstruction.
Electron identification is based on a likelihood discriminant that combines the properties and transition radiation response of the track,
the EM shower shape of the energy cluster and the quality of the track-to-cluster matching.
A \textit{loose} likelihood working point selection~\cite{EGAM-2018-01}, applied in combination with track hit requirements, serves as a preselection requirement and
provides an electron reconstruction and identification efficiency of at least 90\% for isolated electrons with $\pt > 30$~\GeV\
and within the range of 85\%--90\% for isolated electrons with $\pt < 30$~\GeV~\cite{EGAM-2018-01}.\footnote{Electron efficiency values are taken from Run 2, and additional
systematic uncertainties are included to take into account the change in conditions between Run 2 and Run 3, and changes to the MC simulation.}
The energy of electron candidates is calibrated as described in Ref.~\cite{EGAM-2018-01}.
All electron candidates are required to have $\ET>$~7~\GeV\ and be within the acceptance of the ID, $|\eta|<$~2.47.

Muon reconstruction~\cite{PERF-2015-10} in the range of $|\eta|<2.5$ is primarily performed through a global fit to
fully reconstructed tracks in the ID and the MS. Muon candidates are identified
using a \textit{loose}~\cite{PERF-2015-10} identification criterion and required to have $\pt>$~5~\GeV.
The identification criterion has an efficiency of 98\% for isolated muons with $\pt = 5$~\GeV, and rises to 99\% for muons with higher \pt.
At the centre of the detector ($0 < |\eta| < 0.1$), where the MS geometrical coverage is reduced, muon candidates are also identified through the
matching of a fully reconstructed ID track to either an MS track segment or a calorimeter energy deposit consistent with a minimum ionising particle
(calorimeter-tagged muons). In these two cases, the muon candidate's momentum is measured using only the ID track.
For calorimeter-tagged muon candidates, the \pt requirement is raised to $\pt>$~15~\GeV.
No more than one calorimeter-tagged muon is allowed per event.
The momentum of muon candidates is calibrated using the procedure described in Ref.~\cite{PERF-2015-10}.
 
\subsubsection{Primary vertex selection}

The events used in the \hfourl measurement are required to have at least one reconstructed vertex with
at least two associated ID tracks of transverse momentum $\pt>500$~\MeV.
If more than one reconstructed vertex is found, the vertex with the largest $\sum{\pt^2}$
(counting all of its associated tracks) is selected as the primary interaction vertex.
 
The impact parameters of electron and muon candidates are computed relative to the primary interaction
vertex.\footnote{The transverse impact parameter $d_0$ of a charged-particle track is defined in the transverse plane as the distance from the primary vertex to the track's point of closest approach. The longitudinal impact parameter $z_0$ is the distance in the $z$ direction between this point and the primary vertex.}
Muon candidates with transverse impact parameter $d_0$ greater than 1~mm are rejected. 
Both muon and electron candidates are required to have a longitudinal impact parameter $|z_0\sin\theta|$ less than 0.5~mm.

\subsubsection{Event selection}
\label{sec:zz_eventsel}
The rest of the event selection is focused on the selection of a Higgs boson candidate using four lepton candidates.
It is first required that the selected leptons can be matched to at least one of the leptons that triggered the event as described in Section~\ref{sec:data}.
 
Same-flavour opposite-charge (SFOC) lepton candidate pairs are first selected using all electron
and muon candidates in the event that satisfy the preselection.
The SFOC pair with mass closest to the $Z$ boson mass ($m_{Z} = 91.188$~\GeV) is called the leading pair and its mass is labelled \monetwo,
while the other becomes the subleading pair, with mass \mthreefour. If multiple combinations of SFOC pairs can be formed,
a Higgs boson candidate is chosen based on the combination resulting in a leading pair mass \monetwo~closest to the $Z$ boson mass.
If there is more than one type of quadruplet ($4\mu$, $2e2\mu$, $2\mu 2e$ or $4e$) satisfying these selection criteria in the event,
the quadruplet from the channel with highest selection efficiency is chosen as the Higgs boson candidate.
 
At this stage, the three leading lepton candidates of each Higgs boson candidate are required to satisfy $\pt > 20$, 15 and 10~\GeV\ respectively.
Higgs boson candidate events are subjected to further selection requirements on the SFOC pair masses, the lepton candidate separation, a $J/\psi$ mass veto,
the impact parameter significance ($d_0/\sigma(d_0)$) of the lepton candidates, and the quality of the vertex of the event, as outlined in Table~\ref{tab:evtSel}.
 
In addition, track- and calorimeter-based isolation requirements are imposed on the lepton candidates to suppress contributions from the
$t\bar{t}$ and \zjets\ reducible backgrounds. A track-based isolation requirement is defined using the scalar \pt sum of all tracks with $\pt > 500$~\MeV\ that lie
within a cone of $\Delta R = 0.3$ around the muon or electron candidate
and that  either originate from the primary vertex or
have $|z_0 sin \theta| < 3$~mm. For lepton candidates with $\pt > 33$~\GeV, the size of this cone falls linearly as a function of \pt to a minimum cone size of 0.2 at 50~\GeV.
Similarly, a calorimeter-based isolation requirement
based on the scalar \ET sum is calculated from the positive-energy topological clusters that are not associated with the lepton candidate's
track within a cone of $\Delta R = 0.2$ around the lepton candidate. This calorimeter-based isolation is corrected for electron
shower leakage, and pile-up and UE contributions. Both the track- and calorimeter-based isolation are corrected for
track and topological-cluster contributions from the other lepton candidates.
For muons, the sum of the track isolation and 0.4 times the value of the calorimeter isolation is required to be less than 16\% of the lepton candidate's \pt.
For electrons, the sum of the track isolation is required to be less than 15\% of the lepton candidate's \pt and the calorimeter isolation is required to be less than 20\% of the lepton candidate's \pt.

If an extra prompt lepton candidate with $\pt > 12$~\GeV\ and passing all identification and isolation requirements detailed above
is present in the event, the quadruplet of lepton candidates selected as the Higgs boson candidate is chosen using a method based on the ME\@.
The ME for the Higgs boson decay is calculated at LO using \progname{MadGraph5\_aMC@NLO} for each quadruplet in the event,
and the quadruplet with the highest ME value is chosen, with the reconstructed lepton momentum vectors used as inputs to the calculation. This procedure increases the probability of selecting the correct
Higgs boson candidate in cases where the extra lepton comes from the decay of a vector boson or top quark in \VH-leptonic or \ttH/$tH$ production.
 
The quadruplet defined at this stage of the selection is determined to be the final Higgs boson candidate.
 
The four-lepton mass resolution (in particular the long radiative mass tail) is improved by accounting for reconstructed final-state
radiation (FSR) photons originating from the $Z$ boson decay products, as described in Refs.~\cite{HIGG-2018-28, HIGG-2018-29}.
The Higgs boson candidate is required to have a four-lepton mass of $105 < m_{4\ell} < 160$~\GeV.
 
The selection efficiency, relative to the full phase space, is estimated from simulated events to be 27\%, 23\%,
14\% and 13\% for the $4\mu$, $2e2\mu$, $2\mu2e$ and $4e$ final states, respectively.
 
\begin{table}
\centering
\caption{Summary of the event selection requirements for leptons and Higgs boson candidates outlined in Section~\ref{sec:zz_reconstruction}. SFOC lepton pairs are same-flavour opposite-charge lepton pairs. For the mass requirement of the subleading lepton pair, $m_{\textrm{threshold}}$ is 12~\GeV\ for $m_{4\ell}<140$~\GeV, and rises linearly to 50~\GeV\ for $m_{4\ell} = 190$~\GeV.   \label{tab:evtSel} }
\vspace{0.2cm}
\begin{tabular}{ll}
\toprule
\multicolumn{2}{c}{\textbf{Leptons}} \\
\hline
Muons & $\pt>5$~\GeV, $\abseta< 2.5$  \\
Electrons & $\et>7$~\GeV, $\abseta<2.47$  \\
\hline
\multicolumn{2}{c}{\textbf{Lepton selection and pairing}} \\
\hline
Lepton kinematics & $\pt > 20, 15, 10$~\GeV  \\
Leading pair ($m_{12}$) & SFOC lepton pair with smallest $|m_Z-m_{\ell\ell}|$  \\
Subleading pair ($m_{34}$) & remaining SFOC lepton pair with  smallest $|m_Z-m_{\ell\ell}|$  \\
\hline
\multicolumn{2}{c}{\textbf{Event selection (at most one Higgs boson candidate per channel)}} \\
\hline
Mass requirements & 50~\GeV$ < m_{12} < 106$~\GeV \, and  $m_{\textrm{threshold}} < m_{34} < 115$~\GeV  \\
Lepton separation & $\Delta R(\ell_{i},\ell_{j}) > 0.1$  \\
 
$J/\psi$~veto & $m(\ell_{i},\ell_{j}) > 5$~\GeV\  for all SFOC lepton pairs  \\
Impact parameter & $|d_{0}|/\sigma(d_{0}) < $5 (3) for electrons (muons) \\
Mass window &  $105~\GeV < m_{4\ell} < 160$~\GeV  \\
Vertex selection & $\chi^{2}/N_{\mathrm{dof}} <$ 6 (9) for $4\mu$ (other channels) \\
If extra lepton with $\pt > 12$~\GeV\ & quadruplet with largest matrix element value \\
\bottomrule
\end{tabular}
\end{table}


\subsubsection{Fiducial region}
\label{sec:zz_fiducial}
 
The fiducial region is defined using simulation at particle level and the selection requirements outlined in
Table~\ref{tab:FiducialRegion}. To minimise model-dependent acceptance extrapolations, these are chosen to
closely match the selection requirements outlined in Sections~\ref{sec:zz_emureco}-\ref{sec:zz_eventsel}.
 
The fiducial selection is applied to final-state
electrons and muons that do not originate from hadrons or $\tau$-lepton decays, after `dressing', where the
four-momenta of photons within a cone of size $\Delta R = 0.1$ around the lepton are added to the
lepton's four-momentum. Photons that originate from hadron decays are excluded from this procedure.
 
The quadruplet selection using the selected dressed leptons follows the same procedure as in Section~\ref{sec:zz_eventsel}.
In the case of \VH or \STXSttH production, additional leptons not originating from a Higgs boson decay can induce a
`lepton mispairing' when assigning them to the leading and subleading $Z$ bosons. To improve the lepton pairing efficiency,
the ME-based pairing method described in Section~\ref{sec:zz_eventsel} is employed. 
 
The acceptance of the fiducial selection, defined as the ratio of the
number of events passing the particle-level selection to the number of events generated in a given final state  (relative to the full phase space of
$H \rightarrow\ ZZ^* \rightarrow 2\ell 2\ell'$, where $\ell,\ell' = e~\mathrm{or}~\mu$), is about 49\% for each final state.
About  1.4\% of the events that satisfy the detector-level selection fail to satisfy the particle-level selection. This is mostly due to resolution effects.
 
\begin{table}[th]
\centering
\caption{Summary of the event selection requirements that define the fiducial phase space for the \hfourl cross-section measurement. SFOC lepton pairs are same-flavour
opposite-charge lepton pairs. \label{tab:FiducialRegion}}
\vspace{0.2cm}
\begin{tabular}{ll}
\toprule
\multicolumn{2}{c}{\textbf{Leptons}} \\
\hline
Leptons & $\pt>5$ \GeV, $|\eta|<2.7$  \\
\hline
\multicolumn{2}{c}{\textbf{Lepton selection and pairing}} \\
\hline
Lepton kinematics & $\pt > 20, 15, 10 $ \GeV  \\
Leading pair ($m_{12}$) & SFOC lepton pair with smallest $|m_Z-m_{\ell\ell}|$  \\
Subleading pair ($m_{34}$) & remaining SFOC lepton pair with  smallest $|m_Z-m_{\ell\ell}|$  \\
\hline
\multicolumn{2}{c}{\textbf{Event selection (at most one quadruplet per event)}} \\
\hline
Mass requirements & 50 \GeV$ < m_{12} < 106$ \GeV \, and  12 \GeV $< m_{34} < 115$ \GeV  \\
Lepton separation & $\Delta R(\ell_{i},\ell_{j}) > 0.1$ \\
$J/\psi$~veto & $m(\ell_{i},\ell_{j}) > 5$ \GeV \,  for all SFOC lepton pairs  \\
Mass window &  105~\GeV $< m_{4\ell} < 160 $ \GeV  \\
If extra lepton with $\pt>12$~\GeV&  quadruplet with  largest matrix element value \\
\bottomrule
\end{tabular}
 
\end{table}


\label{sec:zz_modeling}


\subsection{Fit procedure}
\label{sec:zz_sig_extraction}
 
The $pp \to H \to ZZ^* \to 4 \ell$ fiducial cross-section is extracted using a procedure analogous to the one presented in Section~\ref{sec:yy_modeling}
for the \hyy measurement.
The signal yield $N_S$ is parameterised as in Eq.~(\ref{eq:NS}) and Eq.~(\ref{eq:XS}), with the fiducial cross-section similarly defined as the product of the total cross-section,
the \hfourl branching ratio and the fiducial acceptance determined by the selection of Section~\ref{sec:zz_fiducial}. The correction factor $\mathcal{C_F}$,
derived from the signal simulated event samples, has values of  55\%, 47\%, 30\% and 26\% for the $4\mu$, $2e2\mu$, $2\mu2e$ and $4e$ final states
respectively. The value of $\mathcal{C_F}$ for the different production modes
ranges from 43\% for \VBF to 29\% for \ttH, and has value of 39\% for the dominant production mode \ggF.
When averaging the four final states and assuming a relative composition of production modes as in the SM, the value of $\mathcal{C_F}$ is 40\%.
 
To extract the number of signal events in each decay final state, invariant mass templates for the Higgs boson signal process and
for the background processes are fitted to a $m_{4\ell}$ binned distribution in data. The signal and \zzstar\ templates are constructed using the simulated event
samples presented in Section~\ref{sec:mcsamples}. 
The normalisations of the signal and non-resonant $ZZ^*$ background are freely floated in a simultaneous fit
to the  $m_{4\ell}$ spectrum within the range of 105--160~\GeV. One single common $ZZ^*$ normalisation factor is used to normalise this background
in the $4\mu$, $2e2\mu$, $2\mu 2e$ and $4e$ final states. Reducible backgrounds, mainly consisting of the \zjets, \ttbar and $WZ$ processes, are estimated by using
dedicated control regions described in more detail in Section~\ref{sec:zz_bkg} below.
 
The systematic uncertainties detailed in Section~\ref{sec:zz_systematics} are included in the fit and implemented as nuisance parameters.
They include, among other things, variations of the signal and background template shapes, and variations of the reducible background normalisation.


\subsection{Reducible background estimation}
\label{sec:zz_bkg}

Background contributions from processes such as \zjets, \ttbar and $WZ$ can satisfy the event selection due to the presence of at least one jet, one photon or one lepton from a hadron decay
that is misidentified as a prompt lepton. These reducible backgrounds are significantly smaller than the non-resonant $ZZ^*$ background and are estimated by using data where possible with different approaches as described in Refs.~\cite{HIGG-2018-28,HIGG-2018-29} and outlined below. Due to the comparatively higher rate of misidentification at lower momentum, the data driven approaches are split into $\ell\ell+\mu\mu$ and $\ell\ell+ee$ final states, which refer to the channels where the sub-leading lepton pair consists of muons or electrons respectively.
All the control regions directly used to estimate the reducible backgrounds are mutually exclusive to each other and to the signal regions.
 
In the \llmumu\ final states, the normalisations of the \zjets\ and \ttbar\ backgrounds are determined by performing fits to the invariant mass spectrum
of the leading lepton pair in dedicated independent control regions that target each of the two background processes for each final state. 
Depending on the background process being targeted, the control regions are formed by relaxing the $\chi^2$ requirement on the four-lepton vertex fit, and
by inverting or relaxing isolation and/or impact-parameter requirements on the subleading muon candidate pair.
Additional control regions ($e\mu \mu\mu$ and \llmumuSS) are used to improve the background yield estimate by reducing the statistical uncertainty
in the fitted normalisation.
Transfer factors to extrapolate from the control regions to the signal region are obtained
separately for \ttbar\ and \zjets\ using simulation. The $m_{4\ell}$ shape for the two processes in each final state is obtained from simulation.
 
For the \llee\ final states, an \llee\ control region selection requires the electrons in the subleading lepton candidate pair to have the same charge,
and relaxes the identification, impact parameter and isolation requirements on the electron candidate with the lowest \ET.
This electron candidate, denoted by $X$, can be a light-flavour jet, an electron from a photon conversion or an electron from a heavy-flavour hadron decay.
The heavy-flavour background, which accounts for about 35\% of the reducible background in these channels, is determined from simulation, whereas the light-flavour and photon-conversion backgrounds are
obtained using the sPlot~method~\cite{splot}. This method is based on a fit to the number of hits in the innermost ID layer, performed in the data control region.
Transfer factors to extrapolate from the
\llee\ control region to the signal region for the light-flavour jets and converted photons are
obtained from simulated event samples, and are corrected using a $Z+X$ data control region.
The $m_{4\ell}$ shape is taken from the control region for the light-flavour jets and converted-photons components and from simulation for the heavy-flavour background.
The agreement between simulation and data of the reducible background mass shape is checked in a region where the selection criteria on the $d_0$
and isolation are relaxed~\cite{HIGG-2018-28,HIGG-2018-29} and no significant discrepancies are found.
 
Additional contributions from rare processes, such as $t\bar{t}Z$, $t\bar{t}W$ and $VVV$ are estimated from simulation.


\subsection{Systematic uncertainties}
\label{sec:zz_systematics}

 
The \hfourl fiducial cross-section measurement is affected by both experimental and theoretical uncertainties.
Experimental uncertainties include uncertainties in the efficiency of electron and muon reconstruction, identification,
isolation and trigger selections, and in electron and muon energy calibration.
Theoretical uncertainties include the modelling of the signal and background processes.
 
\subsubsection{Experimental uncertainties}
\label{subsec:zz_systematics_exp}

Uncertainties in the electron (muon) reconstruction, identification, energy-momentum scale and resolution, and isolation efficiency have an impact on the
expected signal yield and on the shape of the signal and $ZZ^*$ background. These uncertainties have an impact of approximately $6.3 \%$ ($3.8\%$) on the measurement.
The electron (muon) uncertainties are assessed using Run 2 (Run 3) data and simulation with methods described in Refs.~\cite{EGAM-2018-01,PERF-2017-03, PERF-2015-10}.
For electrons, additional uncertainties are included to take into account the change in conditions between Run 2 and Run 3, and changes to the MC simulation.
Lepton trigger efficiency uncertainties have a negligible impact.
 
The impact of the precision of the Higgs boson mass measurement of $m_{H} = 125.09 \pm 0.24$~\GeV~\cite{HIGG-2014-14} is also negligible.
 
The uncertainty in the predicted yields due to uncertainties related to the pile-up modelling is below 1\%.
 
The uncertainty in the luminosity measurement impacts the expected yields of both the signal
and background contributions, except for background estimates in which the data-driven methods can reduce
its impact. As discussed in Section~\ref{sec:data}, the luminosity uncertainty is 2.2\%.

For the data-driven measurement of the reducible background, three sources of uncertainty are considered following the
strategy described in~Ref.~\cite{HIGG-2018-29}: a statistical uncertainty, an overall normalisation systematic uncertainty for each of
\llmumu\ and \llee, and a shape systematic uncertainty that varies for each final state.
The joint impact of these sources of uncertainty on the cross-section measurement is below 1\%.

\subsubsection{Theoretical uncertainties}
\label{subsec:zz_systematics_theory_couplings}
 
Sources of theoretical uncertainty include the effects of missing higher-order corrections, PS and UE modelling,
and PDF plus $\alpha_\mathrm{s}$ uncertainties. They affect the modelling of both signal and background processes.
 
For the signal, the sources of theoretical uncertainty are the same as those described in Section~\ref{sec:yy_systematics} for the \hyy measurement.
For the fiducial cross-section measurement, they impact the expected signal yield
through their effect on the correction factor $\mathcal{C_F}$, which includes the effects of the selection efficiency and of out-of-acceptance corrections.
 
Uncertainties due to missing higher-order QCD effects for the signal are estimated by using the same scheme as in Refs.~\cite{bendavid2018les, HIGG-2018-30}.
These uncertainties have a negligible impact on the fiducial cross-section measurement.

The effects of PS and UE modelling uncertainties are estimated by using tune eigenvector
variations and comparisons between acceptances calculated with \progname{Pythia 8} and \progname{Herwig 7} PS algorithms.
This uncertainty has a negligible impact on the fiducial cross-section measurement.

The impact of the PDF uncertainty is estimated by using the 30 eigenvector variations of the \progname{PDF4LHC\_NLO\_30}
Hessian PDF set calculated at NLO, following the \progname{PDF4LHC} recommendations~\cite{Butterworth:2015oua}.
This uncertainty has a negligible impact on the fiducial cross-section measurement.

Finally, additional uncertainties in $\mathcal{C_F}$ arise from uncertainties in the relative production mode composition,
which can introduce a bias in the unfolding method.
The size of the uncertainty is assessed by varying the production cross-sections within
their measured uncertainties, as taken from Ref.~\cite{HIGG-2016-22}, with a resulting impact of less than 1\%.
 
When extrapolating to the full phase space to derive the total cross-section measurement the impact of these sources of uncertainty on the acceptance is
also included, alongside an uncertainty in the \hfourl decay branching ratio of 2.2\%~\cite{Bredenstein:2006rh,Bredenstein:2006ha}.
At this stage the PS and UE modelling uncertainties are no longer negligible, and have an impact of 1.4\%.

The \zzstar\ process normalisation is derived from a simultaneous fit to the signal region and to sideband regions, hence the theoretical uncertainties
in the \zzstar\ normalisation such as from missing higher-order effects in QCD, PDF uncertainties and PS modelling do not have any impact on this measurement.
In addition, the $m_{4\ell}$ shape obtained from \progname{Sherpa} is compared with that obtained from \progname{MadGraph5\_aMC@NLO},
and the difference is taken as an additional source of systematic uncertainty.
The difference between \progname{Sherpa} and \progname{MadGraph5\_aMC@NLO} on the predicted $m_{4\ell}$ shape varies linearly from approximately $\mp 3\%$ at
low $m_{4\ell}$ to $\pm 3\%$ at high $m_{4\ell}$. These sources of uncertainty have an impact of about 1\% on the cross-section measurement.

The uncertainty in the gluon-induced \zzstar\ process is taken into account by changing the relative composition
of the quark-initiated and gluon-initiated \zzstar\ components according to the theoretical uncertainty in their respective predicted cross-sections. This uncertainty has a negligible impact on the fiducial cross-section measurement.


\subsection{Results}
\label{sec:zz_results}

The observed number of events in each of the four decay final states ($4\mu$, $2e2\mu$, $2\mu 2e$ and $4e$), and the expected signal and background yields after the fit to the data (post-fit), are presented in Table~\ref{tab:recoyields}.
For illustrative purposes, the events shown in this table are selected in a narrower mass window ($115 < \mfourl < 130$~\GeV) relative to that
used in the fit to the data.
 
Figure~\ref{fig:recoyields} shows the expected (post-fit) and observed (data) four-lepton invariant mass distribution.
 
\begin{table*}[htbb!]
\centering
\caption{Predicted SM signal (pre-fit), expected (post-fit) and observed numbers of events in the four decay final states after the event selection, in the mass range \hbox{115~\gev$< \mfourl <$ 130~\gev}.
The sum of the post-fit Higgs boson events and the estimated background yields (reported as Total) are compared with the data. Combined statistical and systematic
uncertainties are included for the predictions (see Section~\ref{sec:zz_systematics}).
\label{tab:recoyields}}
\vspace{0.2cm}
 
\begin{tabular}{wc{2\mylen}
wr{\mylen}@{$\,\pm\,$}wl{\mylen}
wr{\mylen}@{$\,\pm\,$}wl{\mylen}
wr{\mylen}@{$\,\pm\,$}wl{\mylen}
wr{\mylen}@{$\,\pm\,$}wl{\mylen}
wr{\mylen}@{$\,\pm\,$}wl{\mylen}
wc{2\mylen}}
\toprule
Final &  \multicolumn{2}{c}{Signal} & \multicolumn{2}{c}{Signal} & \multicolumn{2}{c}{\zzstar} & \multicolumn{2}{c}{Other} & \multicolumn{2}{c}{\multirow{2}{*}{Total}} & \multirow{2}{*}{Observed} \\
state & \multicolumn{2}{c}{SM (pre-fit)} & \multicolumn{2}{c}{(post-fit)} & \multicolumn{2}{c}{background} & \multicolumn{2}{c}{backgrounds} & \multicolumn{2}{c}{} &  \\
\hline
$4\mu$ & 14.8 & 1.0 & 11.3 & 0.8 & 8.3 & 0.6 & 1.0 & 0.3 & 20.6 & 1.0 & 23 \\
$2e2\mu$ & 11.1 & 0.8 & 8.5 & 0.6 & 6.5 & 0.4 & 1.0 & 0.3 & 16.0 & 0.8 & 13 \\
$2\mu2e$ & 7.0 & 1.3 & 5.4 & 1.0 & 3.2 & 0.6 & 0.9 & 0.1 & 9.4 & 1.2 & 12 \\
$4e$ & 7.4 & 1.5 & 5.7 & 1.1 & 3.1 & 0.7 & 0.8 & 0.2 & 9.6 & 1.4 & \phantom{1}9 \\
\noalign{\vspace{0.05cm}}
\hline
Total & 40.3 & 3.8 & 30.9 & 2.9 & 21.1 & 2.0 & 3.6 & 0.7 & 55.6 & 4.4 & 57 \\
 
\bottomrule
\end{tabular}
 
\end{table*}

\begin{figure}[!htbp]
\centering
\includegraphics[width=0.55\linewidth]{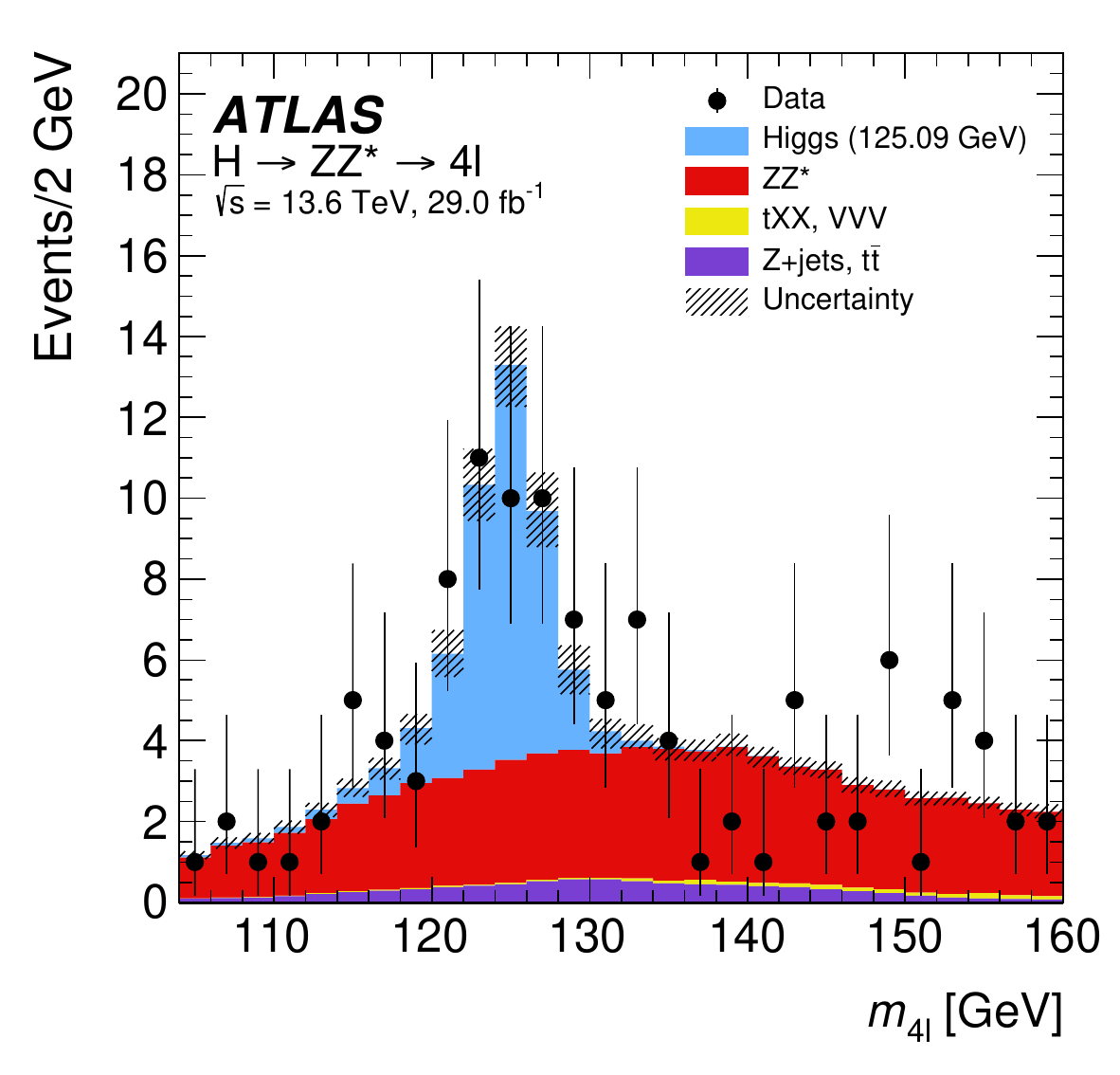}
 
\caption{The observed and expected (post-fit) inclusive four-lepton invariant mass distribution for the selected Higgs boson candidates, shown
for an integrated luminosity of \Lum\ at $\sqrt{\mathrm{s}}=$~13.6~\TeV.
The uncertainty in the prediction is shown by the hatched band, and includes the theoretical uncertainties in the SM cross-section for
the signal and the main background processes. The $tXX$ label indicates the sum of the $t\bar{t}W$ and $t\bar{t}Z$ processes.
}
\label{fig:recoyields}
\end{figure}

The value of the fiducial cross-section, extracted from the  \mfourl\ template fit described in Section~\ref{sec:zz_sig_extraction}, is measured to be \FiducialZZ\
and is in agreement with the SM prediction of \FiducialZZExp.  
The breakdown of the total uncertainty into its different components, derived with the procedure described in Section~\ref{sec:yy_results}, is detailed in Table~\ref{tab:syst_summary_h4l}.
 
\begin{table}[htbp]
\centering
\caption{Breakdown of the relative uncertainties in the fiducial \hfourl cross-section measurement.}
\begin{tabular}{lS[table-format=< 2.1,table-align-comparator = false]}
\toprule
Source  &  {Uncertainty [\%]} \\
\hline
Statistical uncertainty  & 25.1 \\
Systematic uncertainty   & 7.9 \\
\quad Electron uncertainties & 6.3 \\
\quad Muon uncertainties & 3.8 \\
\quad Luminosity & 2.2 \\
\quad \zzstar\ theoretical uncertainties & 0.7 \\
\quad Reducible background estimation & 0.6 \\
\quad Other uncertainties & < 1.0 \\
\hline
Total & 26.4\\
\bottomrule
\end{tabular}
\label{tab:syst_summary_h4l}
\end{table}


\section{Total cross-section measurements}
\label{sec:comb_results}

Assuming SM values for the fiducial acceptances and for the branching fractions of the two channels, the fiducial measurements are extrapolated to the full phase space.
When performing the extrapolation to the full phase space, additional uncertainties in the acceptance and in the branching fraction are considered
(see Sections~\ref{sec:yy_systematics} and ~\ref{sec:zz_systematics}).
The total Higgs boson production cross-section at 13.6~\TeV\ is measured to be \TotalYYShort\ using the $H\to \gamma \gamma$ channel and
\TotalZZShort\ using the $H\to 4\ell$ channel. The two measurements are compatible with a p-value of 20\%.
 
A likelihood combination of the two decay channels is performed, following the method described in Ref.~\cite{HIGG-2017-11}.
 
\begin{figure}[htbp]
\centering
\includegraphics[width=0.7\textwidth]{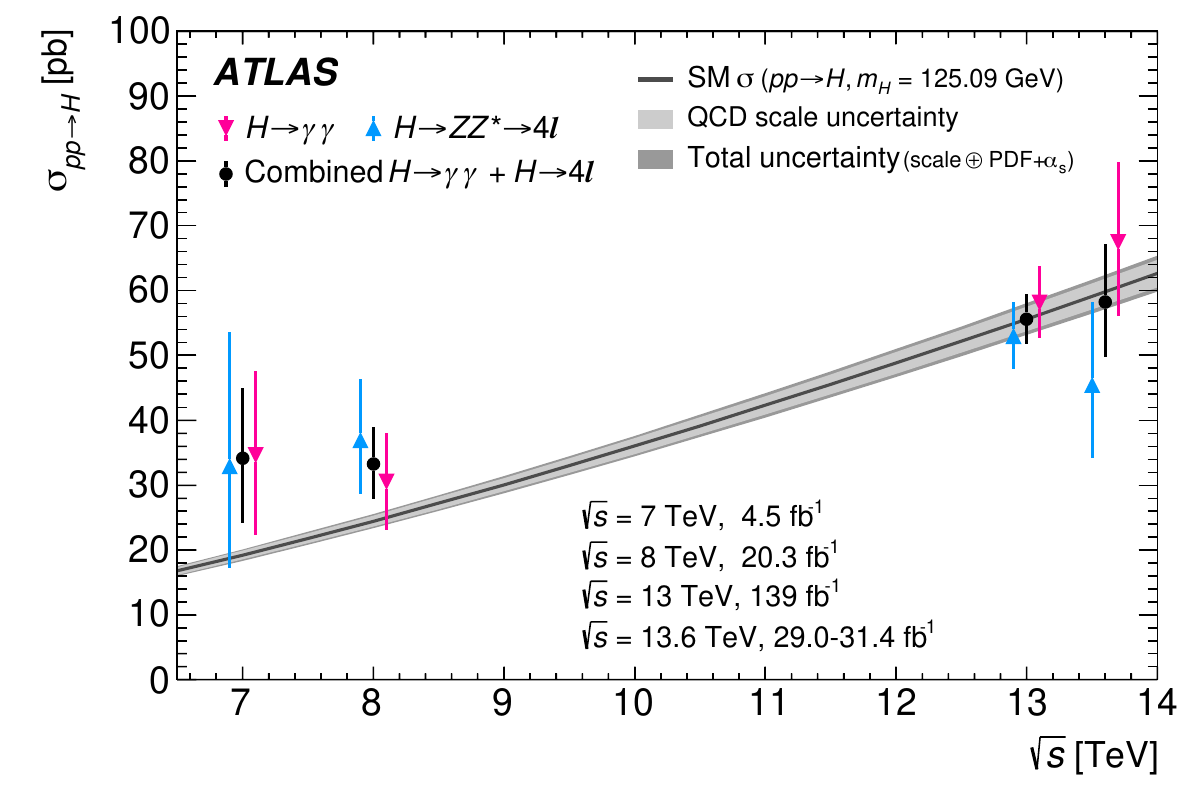}
\caption{ Values of the $\sigma(pp \to H)$ measurements from this and previous~\cite{HIGG-2014-11,HIGG-2022-04} ATLAS publications as a
function of the \pp centre-of-mass energy.
The SM predicted values and their uncertainties are shown by the shaded band.  The individual channel results are offset along the $x$-axis for display purposes.
}
\label{fig:xs}
\end{figure}

Experimental and theoretical uncertainties that affect both channels are correlated via common nuisance parameters.
The correlated experimental uncertainties include the uncertainties in the integrated luminosity, in the description of pile-up in the simulation,
in the common electron--photon energy scale, in the Higgs boson mass value, and in the relative contributions of the
different Higgs boson production modes.
Additionally, the common sources of theoretical uncertainty in the $\hfourl$ and $\hgg$ branching fractions ($\alpha_\mathrm{s}$, $b$- and $c$-quark masses, and partial decay widths into the main decay channels, such as two vector bosons, two gluons, or a $b\bar{b}$ pair) are also correlated.
Finally, the theoretical uncertainties in the acceptance factor due to missing higher-order QCD effects, PDF variations,
variations of the modelling of the PS, and signal composition are also correlated.
 
The asymptotic approximation~\cite{Cowan:2010st} for the distribution of the profile likelihood ratio is assumed in the computation of uncertainties.
The validity of this approximation was verified in previous analyses by performing pseudo-experiments.
 
The total Higgs boson production cross-section, obtained by combining the \hyy and \hfourl results, is \TotalCOMBLong\ at 13.6~\TeV. All three results (\hyy, \hfourl and their combination)
are in agreement with the SM prediction of \TotalCOMBExp.
The nuisance parameters associated with the position of the signal mass peak do not show any significant pull.
 
The values of the total cross-section determined from this analysis, and those from previously published ATLAS studies~\cite{HIGG-2014-11,HIGG-2022-04},
are shown in Figure~\ref{fig:xs} as a function of the \pp centre-of-mass energy.
The measurements at the new centre-of-mass energy of 13.6 \TeV\ are in good agreement with the SM prediction.


\FloatBarrier
\section{Conclusion}
\label{sec:conclusion}

The $pp$ collision data recorded with the ATLAS detector at $\sqrt{s} = 13.6$~\TeV
are used to derive the first measurement of the \hyy and \hzz cross-sections at this new LHC centre-of-mass energy,
with corresponding integrated luminosities of 31.4 and 29.0~\ifb, respectively.
The cross-section measurements are restricted to kinematic phase spaces of the Higgs boson decay
products that closely match the selection criteria applied at detector level,
and are corrected for detector effects.
The measured fiducial cross-sections are \FiducialYYShort\ for the \hyy\ channel and \FiducialZZShort\ for the \hfourl\ channels. They are in agreement
with the corresponding Standard Model predictions of \FiducialYYExpNum\ and \FiducialZZExpNum.
 
Assuming SM values for the acceptances and the branching fractions of the two channels, the fiducial measurements are extrapolated to the
full phase space, yielding total Higgs boson production cross-sections \TotalYYShort\ and  \TotalZZShort\ at 13.6~\TeV for the \hyy and \hzz channels,
respectively. These measurements are combined into a measurement of \TotalCOMBShort, in agreement with the SM prediction of \TotalCOMBExp.


\section*{Acknowledgements}


We thank CERN for the very successful operation of the LHC, as well as the
support staff from our institutions without whom ATLAS could not be
operated efficiently.
 
We acknowledge the support of
ANPCyT, Argentina;
YerPhI, Armenia;
ARC, Australia;
BMWFW and FWF, Austria;
ANAS, Azerbaijan;
CNPq and FAPESP, Brazil;
NSERC, NRC and CFI, Canada;
CERN;
ANID, Chile;
CAS, MOST and NSFC, China;
Minciencias, Colombia;
MEYS CR, Czech Republic;
DNRF and DNSRC, Denmark;
IN2P3-CNRS and CEA-DRF/IRFU, France;
SRNSFG, Georgia;
BMBF, HGF and MPG, Germany;
GSRI, Greece;
RGC and Hong Kong SAR, China;
ISF and Benoziyo Center, Israel;
INFN, Italy;
MEXT and JSPS, Japan;
CNRST, Morocco;
NWO, Netherlands;
RCN, Norway;
MEiN, Poland;
FCT, Portugal;
MNE/IFA, Romania;
MESTD, Serbia;
MSSR, Slovakia;
ARRS and MIZ\v{S}, Slovenia;
DSI/NRF, South Africa;
MICINN, Spain;
SRC and Wallenberg Foundation, Sweden;
SERI, SNSF and Cantons of Bern and Geneva, Switzerland;
MOST, Taiwan;
TENMAK, T\"urkiye;
STFC, United Kingdom;
DOE and NSF, United States of America.
In addition, individual groups and members have received support from
BCKDF, CANARIE, Compute Canada and CRC, Canada;
PRIMUS 21/SCI/017 and UNCE SCI/013, Czech Republic;
COST, ERC, ERDF, Horizon 2020 and Marie Sk{\l}odowska-Curie Actions, European Union;
Investissements d'Avenir Labex, Investissements d'Avenir Idex and ANR, France;
DFG and AvH Foundation, Germany;
Herakleitos, Thales and Aristeia programmes co-financed by EU-ESF and the Greek NSRF, Greece;
BSF-NSF and MINERVA, Israel;
Norwegian Financial Mechanism 2014-2021, Norway;
NCN and NAWA, Poland;
La Caixa Banking Foundation, CERCA Programme Generalitat de Catalunya and PROMETEO and GenT Programmes Generalitat Valenciana, Spain;
G\"{o}ran Gustafssons Stiftelse, Sweden;
The Royal Society and Leverhulme Trust, United Kingdom.
 
The crucial computing support from all WLCG partners is acknowledged gratefully, in particular from CERN, the ATLAS Tier-1 facilities at TRIUMF (Canada), NDGF (Denmark, Norway, Sweden), CC-IN2P3 (France), KIT/GridKA (Germany), INFN-CNAF (Italy), NL-T1 (Netherlands), PIC (Spain), ASGC (Taiwan), RAL (UK) and BNL (USA), the Tier-2 facilities worldwide and large non-WLCG resource providers. Major contributors of computing resources are listed in Ref.~\cite{ATL-SOFT-PUB-2021-003}.


\printbibliography
 
\clearpage
 
\begin{flushleft}
\hypersetup{urlcolor=black}
{\Large The ATLAS Collaboration}

\bigskip

\AtlasOrcid[0000-0002-6665-4934]{G.~Aad}$^\textrm{\scriptsize 102}$,
\AtlasOrcid[0000-0002-5888-2734]{B.~Abbott}$^\textrm{\scriptsize 120}$,
\AtlasOrcid[0000-0002-1002-1652]{K.~Abeling}$^\textrm{\scriptsize 55}$,
\AtlasOrcid[0000-0001-5763-2760]{N.J.~Abicht}$^\textrm{\scriptsize 49}$,
\AtlasOrcid[0000-0002-8496-9294]{S.H.~Abidi}$^\textrm{\scriptsize 29}$,
\AtlasOrcid[0000-0002-9987-2292]{A.~Aboulhorma}$^\textrm{\scriptsize 35e}$,
\AtlasOrcid[0000-0001-5329-6640]{H.~Abramowicz}$^\textrm{\scriptsize 151}$,
\AtlasOrcid[0000-0002-1599-2896]{H.~Abreu}$^\textrm{\scriptsize 150}$,
\AtlasOrcid[0000-0003-0403-3697]{Y.~Abulaiti}$^\textrm{\scriptsize 117}$,
\AtlasOrcid[0000-0002-8588-9157]{B.S.~Acharya}$^\textrm{\scriptsize 69a,69b,m}$
% ...
\bigskip
\\

% ...
$^{29}$Physics Department, Brookhaven National Laboratory, Upton NY; United States of America.\\
% ...
$^{35}$$^{(a)}$Facult\'e des Sciences Ain Chock, R\'eseau Universitaire de Physique des Hautes Energies - Universit\'e Hassan II, Casablanca;$^{(b)}$Facult\'{e} des Sciences, Universit\'{e} Ibn-Tofail, K\'{e}nitra;$^{(c)}$Facult\'e des Sciences Semlalia, Universit\'e Cadi Ayyad, LPHEA-Marrakech;$^{(d)}$LPMR, Facult\'e des Sciences, Universit\'e Mohamed Premier, Oujda;$^{(e)}$Facult\'e des sciences, Universit\'e Mohammed V, Rabat;$^{(f)}$Institute of Applied Physics, Mohammed VI Polytechnic University, Ben Guerir; Morocco.\\
% ...
$^{49}$Fakult\"{a}t Physik , Technische Universit{\"a}t Dortmund, Dortmund; Germany.\\
% ...
$^{55}$II. Physikalisches Institut, Georg-August-Universit\"{a}t G\"ottingen, G\"ottingen; Germany.\\
% ...
$^{69}$$^{(a)}$INFN Gruppo Collegato di Udine, Sezione di Trieste, Udine;$^{(b)}$ICTP, Trieste;$^{(c)}$Dipartimento Politecnico di Ingegneria e Architettura, Universit\`a di Udine, Udine; Italy.\\
% ...
$^{102}$CPPM, Aix-Marseille Universit\'e, CNRS/IN2P3, Marseille; France.\\
% ...
$^{117}$Department of Physics, New York University, New York NY; United States of America.\\
% ...
$^{120}$Homer L. Dodge Department of Physics and Astronomy, University of Oklahoma, Norman OK; United States of America.\\
% ...
$^{150}$Department of Physics, Technion, Israel Institute of Technology, Haifa; Israel.\\
$^{151}$Raymond and Beverly Sackler School of Physics and Astronomy, Tel Aviv University, Tel Aviv; Israel.\\
% ...
$^{a}$ Also Affiliated with an institute covered by a cooperation agreement with CERN.\\
$^{b}$ Also at An-Najah National University, Nablus; Palestine.\\
$^{c}$ Also at Borough of Manhattan Community College, City University of New York, New York NY; United States of America.\\
$^{d}$ Also at Center for High Energy Physics, Peking University; China.\\
$^{e}$ Also at Center for Interdisciplinary Research and Innovation (CIRI-AUTH), Thessaloniki; Greece.\\
$^{f}$ Also at Centro Studi e Ricerche Enrico Fermi; Italy.\\
% ...
$^{m}$ Also at Department of Physics, King's College London, London; United Kingdom.\\
% ...

\end{flushleft}


\end{document}